\documentclass[reprint,aip,jcp,superscriptaddress,
floatfix]{revtex4-2}

\usepackage[T1]{fontenc}  
\usepackage[utf8]{inputenc} 
\usepackage{amsmath, amssymb}  
\usepackage[dvipsnames]{xcolor}    
\usepackage{hyperref}  
\hypersetup{  
    colorlinks,  
    linkcolor={RoyalBlue},  
    citecolor={RoyalBlue},  
    urlcolor={RoyalBlue}  
}  
\usepackage{graphicx}  
\usepackage{xspace}
\newcommand{\etal}{\textit{et al}.\xspace}  
\newcommand{\ie}{i.e.,\xspace} 
\newcommand{\eg}{e.g.,\xspace} 

\begin{document}
\title{Identifying sequential residue patterns in bitter and umami peptides}

\author{Arghya Dutta}
\email{argphy@gmail.com}
\altaffiliation{Present address: Institute of Biochemistry II, Faculty of Medicine, Goethe University, Frankfurt, Germany}
\affiliation{Max Planck Institute for Polymer Research, Mainz, Germany}

\author{Tristan Bereau}
\affiliation{Van 't Hoff Institute for Molecular Sciences and Informatics
    Institute, University of Amsterdam, The Netherlands}

\author{Thomas A. Vilgis}
\email{vilgis@mpip-mainz.mpg.de}
\affiliation{Max Planck Institute for Polymer Research, Mainz, Germany}

\begin{abstract}
    The primary structures of peptides, originating from food proteins, affect their taste. Connecting primary structure to taste, however, is difficult because the size of the peptide sequence space increases exponentially with increasing peptide length, while experimentally-labeled data on peptides' tastes remain scarce. We propose a method that coarse-grains the sequence space to reduce its size and systematically identifies the most common coarse-grained residue patterns found in known bitter and umami peptides. We select the optimal patterns by performing extensive out-of-sample tests. The optimal patterns better represent the bitter and umami peptides when compared against baseline peptides, bitter peptides with all hydrophobic residues and umami peptides with all negatively charged residues, and peptides with randomly-chosen residues. Our method complements quantitative structure--activity relationship methods by offering generic, coarse-grained bitter and umami residue patterns that can aid in locating short bitter or umami segments in a protein and in designing new umami peptides.
\end{abstract}

\maketitle

\section{Introduction}

Special compounds trigger specific tastes: sodium chloride (salty), sugars (sweet), acids, phenols, and alkaloids (bitter), glutamic acids and nucleotides (umami). Tastes are crucial because the gustatory system, the sensory system that helps in perceiving taste, often informs us about safe and harmful foods through their tastes\cite{chandrashekar2006receptors}. Further, taste determines most of our food preferences\cite{drewnowski1997taste}. For example, vegetables such as cabbage, cucumber, or spinach, often taste bitter since they contain plant alkaloids---which can be toxic if consumed in large amounts and are known to have excessive bitter taste \cite{drewnowski1997taste, maehashi2009bitter}---and, consequently, we avoid them.

Bitter and umami represent two major taste modalities that peptides commonly have; the third one is sweet\cite{temussi2012good}. Interestingly, while salty, sour, sweet, and bitter were recognized as basic tastes early on, umami was recognized as the fifth basic taste only around the beginning of this century when umami taste receptors were identified \cite{chaudhari2000metabotropic, nelson2002amino, san2005cloning}. As a result, the study of umami peptides is a more recent endeavor compared to, for example, the study of bitter peptides\cite{behrens2011sweet, zhang2017novel}. 

Bitter peptides are often found in fermented foods \cite{matoba1972relationship, lemieux1991bitter} and protein hydrolysates\cite{maehashi2009bitter}, while umami peptides are found in savory foods such as parmesan cheese, fermented soy sauce, and seaweed\cite{behrens2011sweet}. As we tend to avoid bitter foods and seek savory ones, classifying foods based on the taste responses they evoke and modulate, and finding the physicochemical reasons causing those responses are indispensable steps in designing new nutritional and palatable foods. The growing number of curated databases, and analysis tools, of bitter- and umami-tasting foods \cite{dagan2017bitter, dagan2019bitterdb,  rojas2022chemtastesdb} and their relevant taste receptors \cite{huang2016bitterx} indicates recent progress in this direction. Most savory foods result from protein hydrolysis, such as long-time-cooked preparations, fermented foods, \eg soy, fish, oyster sauces, miso pastes, or long-matured and ripened foods, such as cheese or cured meat \cite{yamaguchi2000umami}. It is now well-accepted that amino acids and peptides contribute significantly to the overall taste of such foods \cite{temussi2012good}. Whereas single amino acids are likely to form aroma compounds during thermal and micro-biological processing \cite{van2006formation, zhao2016formation}, peptides remain more stable and contribute significantly to taste. In this paper, we focus on bitter- and umami-tasting peptides as they play the most important role in determining the overall flavor of many foods. 

For example, bitter peptides are well-known to occur in matured cheeses \cite{temussi2012good, toelstede2008sensomics}. They are often produced during cheese ripening because most of the bitter-tasting amino acids are hidden in the caseins. This already suggests a connection between the physicochemical properties of the amino acids and their tastes. Bitter-tasting amino acids are hydrophobic, so it should not be surprising if hydrophobic peptides become abundant and have a strong impact on the flavor of both still-ripening and well-matured cheeses. Well-matured cheeses are also good examples of foods containing umami peptides, similar to soy sauces, miso pastes, ham, cured meat, and matured sausages \cite{zhao2016formation, lioe2010soy, kusumoto2021japanese, heres2021characterization, wang2022effect}.

In fact, more and more taste-relevant peptides are now being discovered in foods from different preparations \cite{juenger2022sensoproteomic}. Thus, the question arises to what extent possible taste qualities of peptides can be identified in advance from their primary structure. In the course of new developments in plant-based foods, these questions are gaining in importance. If, for example, surrogate products are designed based on certain plant proteins, it would be helpful to identify which short sequences of these proteins exhibit particular flavor qualities. These proteins could then be thermally and enzymatically treated to extract flavor peptides, which can then be used as flavor enhancers.  

Taken together, given the tastes of individual amino acids \cite{solms1969taste}, the challenging question is whether the tastes of peptides follow certain residue patterns, determined by their minimum physicochemical properties such as hydrophobicity, polarity, and charge. As already mentioned, hydrophobic amino acids are likely to taste bitter and negatively charged umami, polar ones, as mainly sweet \cite{kawai2012gustatory}.

Traditionally, the study of peptides' tastes relied on the quantitative structure--activity relationships (QSAR) framework that relates peptide descriptors to some desired target property using statistical and machine learning (ML) methods. For peptides, QSAR studies flourished for the additional reason that a peptide's sequential primary structure lends itself easily to developing physicochemical descriptors \cite{hellberg1987peptide, tong2008novel, lin2008new, liang2009using, lee2018machine, xu2019quantitative}. 

For bitter peptides, QSAR has been used along with physicochemical descriptors, for example, to predict threshold concentration for bitterness \cite{kim2006quantitative, soltani2013qsbr}, to predict bitter and non-bitter peptides \cite{rojas2016quantitative, charoenkwan2021ibitterfuse}, to find residue types of bitter di- and tri-peptides\cite{wu2007quantitative}, and to find bioactivity of bitter peptides\cite{yin2010studying}. Predicting taste only based on sequence information has been attempted recently by Charoenkwan \etal \cite{charoenkwan2020ibitter, charoenkwan2021bert4bitter}. Though early on it was conjectured that the positions of residues of a peptide do not affect its taste \cite{matoba1972relationship, ney1979bitterness}, multiple studies since then have found that the residue positions do affect the taste \cite{ishibashi1987bitterness, wu2007quantitative}.

For umami peptides, there are fewer QSAR studies compared to studies done on bitter peptides. One reason for this is that the current experimental methods for measuring umami intensity often fall short in sensitivity and specificity \cite{qi2022research}. Accordingly, defining the target variable for umami intensity proved to be difficult. To bypass this difficulty, the computational methods that have been generally used to predict and analyze umami peptides relied on structural analysis such as homology modeling and molecular docking \cite{dang2014interaction, yu2021identification, wang2022silico, liang2022characteristics, liang2022characterization} of possible umami peptides to umami taste receptors such as T1R1/T1R3 \cite{zhang2008molecular}. Quite recently, physicochemical descriptors \cite{charoenkwan2021umpred} and only sequence information \cite{charoenkwan2020iumami} were used along with ML-based methods to classify umami and non-umami peptides. 

As we have seen so far, QSAR and ML methods focus either on classifying peptides or on predicting values of some target variable using physicochemical descriptors or sequential information. While QSAR methods are generally easy to interpret, the physicochemical descriptors they use come from linear and non-linear dimensionality reduction techniques \cite{bo2021application}---this makes the final models less interpretable. Further, often multiple descriptors are needed to achieve higher prediction accuracy \cite{hellberg1987peptide, venkatarajan2001new, yin2010studying}; this makes the models multi-dimensional and even more difficult to interpret. While ML is shown to make accurate predictions using only sequence information \cite{charoenkwan2021bert4bitter}, they inherit the low interpretability issue often found in ML methods such as deep neural networks. The scarcity of experimentally-verified data on peptides' tastes creates an additional, often rate-limiting, step for black-box ML models that generally works well only when they are trained with a large, labeled dataset.

Thus, while QSAR and ML methods are essential for making accurate predictions given a peptide sequence, their low interpretability becomes an issue if, for example, the aim is to design \emph{new} (\ie out-of-sample) umami peptides or to locate bitter-causing segments in a long protein. Instead, a systematically-derived generic residue pattern, which is possibly connected to the taste, can provide a better starting point and thus can substantially speed up realizing these aims. In this paper, we propose a method that identifies such generic coarse-grained residue patterns that are often found in bitter and umami peptides. The lower granularity of a coarse-grained model is necessary as it helps to find generic residue patterns by reducing the size of the peptide sequence space.

\begin{figure}
    \begin{center}
        \includegraphics[width=\columnwidth]{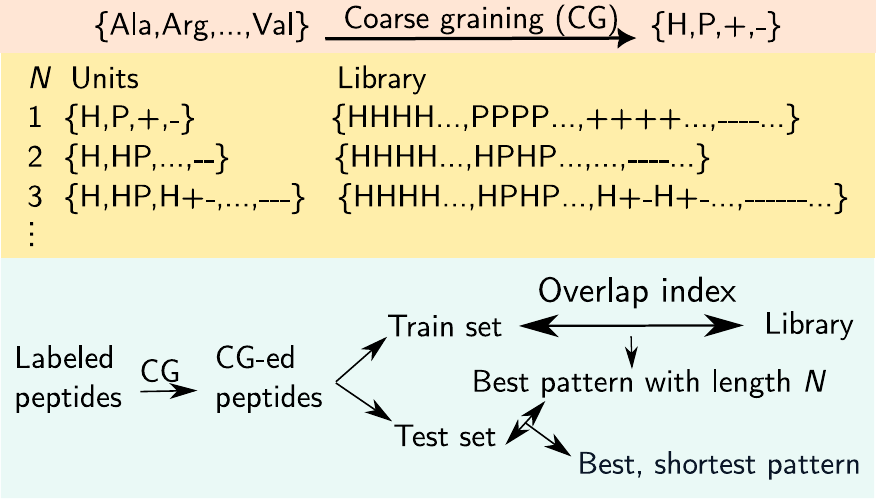}
        \caption{Schematic diagram illustrating the method we used to identify the most common coarse-grained residue patterns present in known bitter and umami peptides.}
        \label{fig:fig1}
    \end{center}
\end{figure}

To this end, we first reduced the size of the peptide sequence space by classifying the amino acids into four coarse-grained residue types: hydrophobic (H), polar and hydrophilic (P), positively charged (+), and negatively charged ($-$) (Fig.~\ref{fig:fig1}). We combinatorially constructed seven comprehensive, increasingly large libraries of peptides with coarse-grained residue patterns. We compiled a database of bitter and umami peptides from the literature. After dividing the database peptides into train and test sets, we compared the library peptides to the coarse-grained bitter and umami peptides from the training sets using a sequence comparison index \cite{schilling2019sequence} and two surrogate measures, defined using the comparison index, for bitterness and umami-ness. This comparison brought out the best residue patterns that have the highest average overlaps with the bitter and the umami peptides, for each library. Finally, we compared the average overlaps of the peptides, constructed from the predicted patterns from different libraries, with peptides from the test sets to find the shortest pattern that has the highest (or close to the highest) overlap. To assess the accuracy of the predicted bitter and umami patterns, we checked if they have higher overlaps with bitter and umami peptides from test sets compared to overlaps with an all-hydrophobic bitter baseline peptide and an all-acidic umami baseline peptide, respectively, and, also, to a peptide with randomly-chosen residues. We used this method to assess the accuracy of our predicted patterns because our goal is to reveal generic residue patterns rather than predicting whether an individual peptide has bitter or umami taste. Taken together, our method systematically expands the currently known set of bitter and umami patterns and suggests a way to identify residue patterns that can be responsible for bitter or umami taste in a peptide or a protein.

\section{Method}

\subsection{Database of labeled peptides}

In this work, we relied on bitter and umami peptides collected from the existing literature. Our principal source is the database of 299 bitter and 140 umami peptides provided by Charoenkwan \etal, who compiled the list from experimentally-validated datasets and literature \cite{charoenkwan2020iumami, charoenkwan2020iumami-data}. To Charoenkwan \etal's list of bitter peptides, we added 24 new bitter peptides that we found in Ney's paper\cite{ney1979bitterness}. As for umami peptides, we added 12 more peptides from the literature\cite{shiyan2021novel, liu2020seven} to Charoenkwan \etal's list. As more than 90\% of the collected peptides are composed of 2--10 amino acid residues, we discarded single residue peptides and peptides with more than 10 residues. This resulted in 292 bitter and 146 umami peptides that were used in this work.

\begin{figure*}
    \begin{center}
        \includegraphics[width=\textwidth]{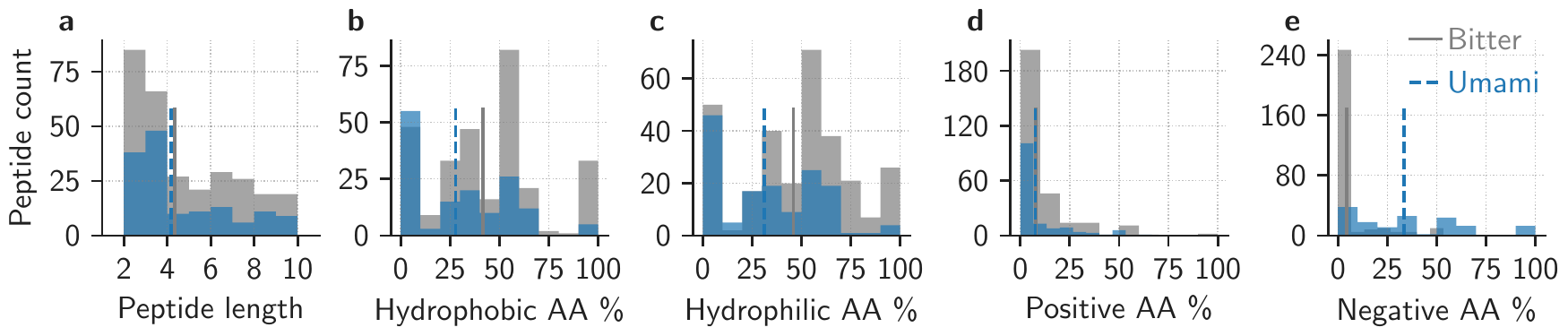}
        \caption{Physicochemical properties of bitter (gray) and umami (blue) peptides from the compiled database. Histograms of (\textbf{a}) peptide lengths and of percentage compositions of amino acid (AA) residue types---(\textbf{b}) hydrophobic, (\textbf{c}) hydrophilic, (\textbf{d}) positively charged, and (\textbf{e}) negatively charged---that comprise the database peptides. While bitter peptides are rich in hydrophobic residues (panel~\textbf{b}), umami peptides mostly contain negative and polar residues (panels~\textbf{e} and ~\textbf{c}). The vertical dashed lines indicate mean values of the distributions.}
        \label{fig:fig2}
    \end{center}
\end{figure*}

\subsection{Coarse-grained representation}

The size of the sequence spaces of peptides, which are made from a combination of the 20 canonical amino acids, grows as $20^n$ where $n$ is the number of residues in the longest allowed peptide. As we have only 292 bitter and 146 umami peptides, we need to reduce the size of the peptide sequence space to make reliable predictions. One way of reducing the size is by coarse-graining the amino acids. We did this by classifying each amino acid into one of the four classes: hydrophobic (H), polar and hydrophilic (P), positively charged (+), or negatively charged ($-$). We have used the Kyte--Doolittle (KD) hydrophobicity scale \cite{kyte1982simple} to find hydrophobic (KD hydrophobicity $>0$) and hydrophilic (KD hydrophobicity $<0$) residues. In this scheme, at physiological pH of 7.4, the 20 canonical amino acids get classified as follows: hydrophobic (H): \{Ala, Cys, Ile, Leu, Met, Phe, Val\}, hydrophilic (P): \{Asn, Gln, Gly, Pro, Ser, Thr, Trp, Tyr\}, positively charged (+): \{Arg, His, Lys\}, and negatively charged ($-$): \{Asp, Glu\}. For example, in our representation scheme, the bitter peptide LLLPGELAK is represented as `HHHPP$-$HH+'. With this representation, the size of the possible peptide sequence space reduces drastically. For example, the number of possible di-peptides reduces from $20^2=400$ to $4^2=16$. We converted all the collected peptides from literature to coarse-grained sequences and then proceeded to construct a library of coarse-grained peptides.

\subsection{Library of coarse-grained peptides}

To extend the prediction beyond the peptide dataset we started with, we need new peptide sequences. To systematically generate new peptide sequences, we constructed seven increasingly larger peptide libraries formed by repeating a fixed set of coarse-grained patterns. While each of these libraries produced two best, \ie most-overlapped, matching patterns for bitter and umami peptides, we also compared patterns from these seven libraries. In this way, we can avoid choosing an unnecessarily large library when a smaller one performs comparably---\ie we will not overfit. We don't seek a very small library of peptides either, as that will lead to underfitting. In the results sections, we will see how until $N=3$ the libraries underfit the data, while beyond $N=5$ the predictive power of the libraries saturates.  

We constructed each library in four steps. First, we fixed the maximum length ($N$) of the repeating patterns. Second, we generated a list of all $\sum_{i=1}^N 4^i$ possible combinatorial patterns containing $N$ or fewer coarse-grained residues. Third, we repeated each pattern with itself to form an arbitrarily long (set to 420 residues in this work) peptide. Finally, we kept only unique full peptides in the final library. For example, in the $N=1$ library, there are only four repeating patterns: \{H, P, +, $-$\} and only four unique peptides: \{HH$\cdots$, PP$\cdots$, ++$\cdots$, and $--\cdots$\}. For $N=3$, the library has $\sum_{i=1}^3 4^i=84$ repeating patterns such as \{H, $-$P, +$-$H,$\cdots$\}, and they can combine to generate 76 unique peptides: \{HH$\cdots$, $-$P$-$P$\cdots$, +$-$H+$-$H$\cdots$, $\cdots$\}. In this way, we went up to $N=7$ and constructed seven libraries. We stopped at $N=7$ because as we increase $N$, the number of peptides in the library increases sharply, and we will risk overfitting the data. For example, the $N=7$ library has $21\,844$ repeating patterns and $21\,736$ unique peptides. Now to compare the library peptides and labeled peptides, we need an index that can measure the similarity between any two coarse-grained sequences.

\subsection{Sequence overlap index}

Following Schilling \etal\cite{schilling2019sequence}, we defined an overlap index between two coarse-grained peptides, X and Y, as the ratio of the number of position-dependent residue matches ($\lvert{\rm X} \cap {\rm Y}\rvert_\textrm{seq}$) and the length of the smaller peptide: 
\begin{equation}
    \label{eq:overlap-index}
    I({\rm X}, {\rm Y}) = \frac{\lvert{\rm X} \cap {\rm Y}\rvert_\textrm{seq}}{\min\left(\lvert {\rm X}\rvert, \lvert {\rm Y}\rvert \right)}.    
\end{equation}
 $\lvert {\rm X}\rvert$ denotes the number of residues in the peptide X. For example, two peptides `$-$H++' and `+H+$--$' have two position-dependent residue matches (`H' and `+' at positions 2 and 3, respectively), so $I$($-$H++, +H+$--$)=2/4=0.5.

The overlap index, $I$, allows us to define a surrogate measure for bitter and umami tastes of a library peptide. For a coarse-grained library peptide, we defined bitterness (umami-ness) as the average overlap between the library peptide and all coarse-grained bitter (umami) peptides from the compiled dataset.

\subsection{Best patterns and their validation}

With this surrogate measure for the bitterness (umami-ness) in our formalism, we considered each of the seven libraries in turn, computed the bitterness (umami-ness) of its constituent peptides, and then sorted them to find five peptides with the largest bitterness (umami-ness) values. (For the $N=1$ library, there are only 4 possible peptides; we chose the best one.) At each sequence position of the repeating patterns of these five peptides, we found the most occurring residue type (from H, P, +, and $-$). By sequentially merging these most occurring residue types, we finally get the best pattern. Note that this composed best pattern has the same length as the maximum length of the repeating patterns ($N$) of a library. Thus, the best pattern depends on the library, the taste type (bitter or umami), and the external database of peptides that we use to measure the taste (bitter- or umami-ness). Therefore choosing a well-curated and sufficiently large (so that the prediction errors are low) set of taste-labeled peptides is crucial in our data-driven approach.

To ensure reproducibility of the predicted pattern, we performed extensive out-of-sample testing. To this end, we split the external taste-labeled peptide database into an 80\% training set and a 20\% test set using stratified random sampling. We used stratified sampling to keep the ratio of bitter and umami peptides roughly similar in the training and the test sets. Otherwise, a random sampling will pick more bitter peptides than umami ones because we have 292 bitter and 146 umami peptides in the peptide database. This, in turn, will result in imbalanced training data for identifying the bitter and umami patterns. Finally, we obtained the best bitter (umami) pattern for each of the seven libraries by using the training set peptides. To gather enough statistics, we repeated this procedure 500 times.

\subsection{Baseline patterns}

Following the literature, we set a peptide with all hydrophobic residues as the baseline bitter peptide\cite{matoba1972relationship, ney1979bitterness, iwaniak2016food} and a peptide with all negatively charged residues as the baseline umami peptide \cite{rhyu2011umami, yu2017structure, qi2022research, wang2022silico}. While the compositions of longer umami peptides are known to be varied \cite{qi2022research}, the importance of the presence of negatively charged acidic residues is generally well accepted in the community. Further, setting a baseline will allow us to quantitatively assess the conjecture that the relative locations of the residues don't affect a peptide's taste \cite{matoba1972relationship, ney1979bitterness}. In Fig.~\ref{fig:fig1}, we have presented the main steps of the complete pipeline that we used in this paper.

\begin{figure*}
    \begin{center}
        \includegraphics[width=\textwidth]{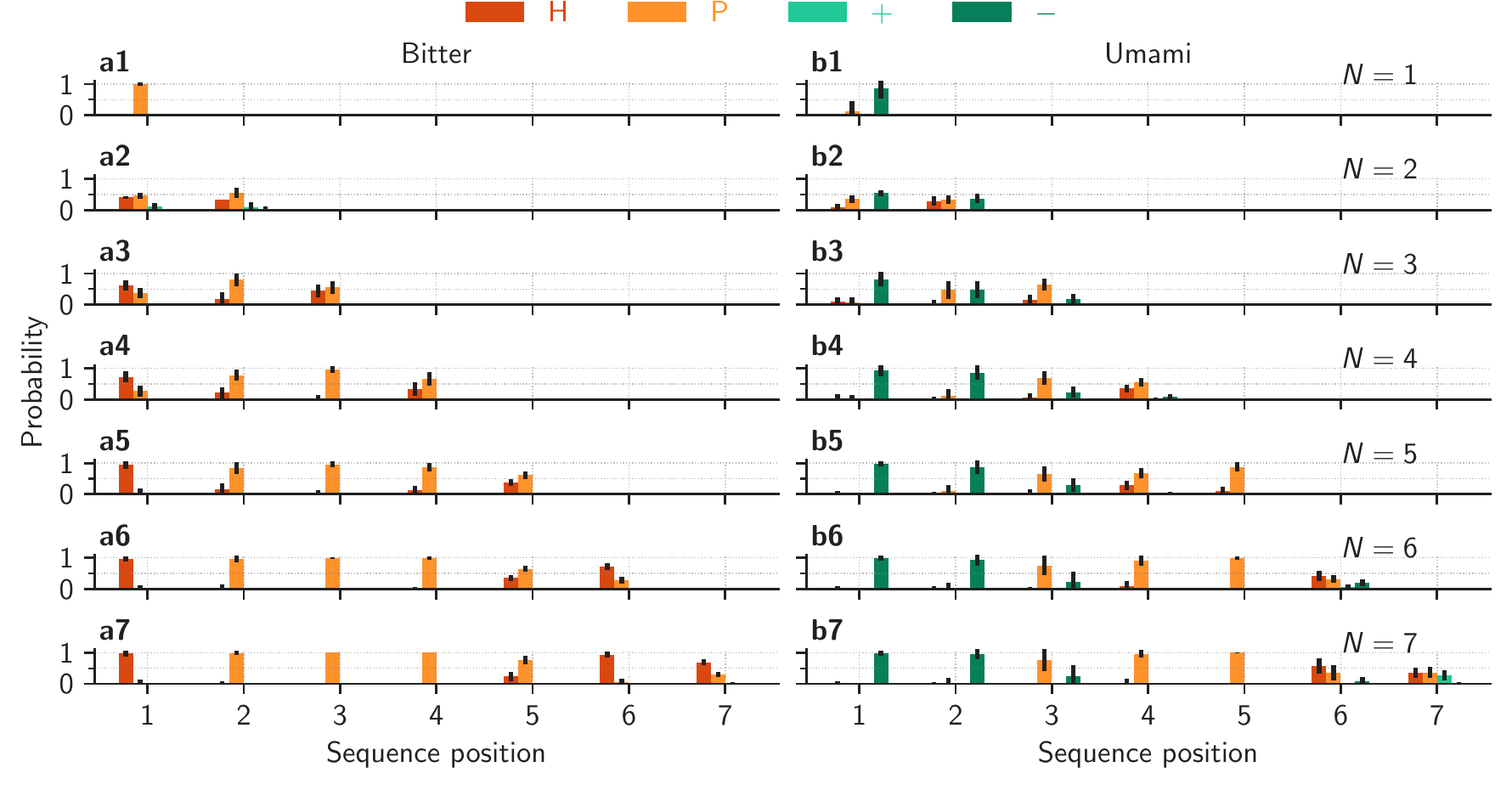}
        \caption{Predicted coarse-grained sequence patterns that have highest overlap with bitter (left panel, \textbf{a1}--\textbf{a7}) and umami (right panel, \textbf{b1}--\textbf{b7}) peptides. The maximum lengths of the peptide library patterns ($N$) increase from top to bottom in each panel. The colored bars show the probabilities of finding a specific residue type at a given sequence position of the predicted pattern. The color codes are displayed at the top. Black error bars indicate standard deviations computed over 500 training sets.}
        \label{fig:fig3}
    \end{center}
\end{figure*}
\section{Results and discussions}

\subsection{Physicochemical properties of the database peptides}

We first analyzed the physicochemical properties of the labeled coarse-grained peptides from the compiled database. Both bitter and umami peptides from the assembled dataset have about 4 residues, on average (Fig.~\ref{fig:fig2}a). Interestingly, $\sim 54\%$ of the database peptides have only two or three residues. This reflects the fact that while consensus often exists regarding the tastes of shorter peptides, there are some disagreements regarding the taste of longer peptides, especially for longer umami peptides \cite{qi2022research}. This makes the exploration of longer peptide patterns even more relevant for the food industry because it can lead to the discovery of new bitter and umami peptides \cite{yu2017structure}.

To find the relative abundance of the four coarse-grained residue types (\ie H, P, +, and $-$), we computed the corresponding histograms (Fig.~\ref{fig:fig2}b--e) of their presence (in \%) in the bitter and umami peptides from the compiled database of peptides. We found that in both bitter and umami peptides, the hydrophilic residues are abundantly present (Fig.~\ref{fig:fig2}c), while positively charged residues are mostly absent (Fig.~\ref{fig:fig2}d). On average, bitter peptides contain more hydrophobic residues than the umami peptides: $\sim 42\%$ compared to $\sim 28\%$ (Fig.~\ref{fig:fig2}b). The umami peptides are richer in negatively charged amino acids ($\sim 33\%$) compared to the bitter peptides ($\sim 5\%$) (Fig.~\ref{fig:fig2}e). Both of these observations are in accord with the current consensus that the hydrophobic residues dominate bitter peptides \cite{ney1979bitterness, iwaniak2016food}, while negative residues dominate umami peptides \cite{rhyu2011umami, yu2017structure, qi2022research, wang2022silico}. We, however, also note the significant presence of hydrophilic residues in bitter peptides. This observation asks for a systematic analysis of residue patterns in the primary sequence of coarse-grained bitter and umami peptides. In the next section, we present our findings from such an analysis.

\subsection{Predicted bitter and umami patterns}

Fig.~\ref{fig:fig3} shows the patterns that best predict bitterness (left panel, Fig.~\ref{fig:fig3}a1--a7) and umami-ness (right panel, Fig.~\ref{fig:fig3}b1--b7) for all seven libraries, from $N=1$ to $N=7$. For $N=1$, we have the smallest library; and the predicted pattern simply picks the most common residue type in bitter and umami peptides from the training sets. For bitter peptides, hydrophilic residues are most common ($\sim 46\%$), followed by hydrophobic residues ($\sim 42\%$) (Fig.~{\ref{fig:fig2}c,b}), while for umami peptides negative residues are most common ($\sim 33\%$), followed by hydrophilic residues ($\sim 31\%$) (Fig.~{\ref{fig:fig2}e,c}). So it is not surprising to find that for the $N=1$ library our algorithm predicts hydrophilic (`P') and negative residues (`$-$') as the best patterns for bitter and umami peptides, respectively (Fig.~\ref{fig:fig3}a1,b1). This prediction, however, is an example of under-fitting the data as we didn't allow for enough complexity (\ie enough peptides) in our library.

The lack of complexity affects the results for the library with $N=2$, too. We find the residues types compete closely at both sequence positions of the predicted patterns (Fig.~\ref{fig:fig3}a2,b2). The algorithm predicts an all hydrophilic residue pattern, `PP', as the best bitter pattern and an all negative residue pattern, `$--$', as the best umami pattern. Though for umami peptides the predicted pattern matches with the literature consensus \cite{qi2022research}, for bitter peptides it does not---similar to the result from the $N=1$ library. The presence of longer peptides with many hydrophilic residues in our bitter peptide dataset subdues the expected `HH' pattern. Interestingly, however, the `HH' pattern does get predicted as the second best bitter pattern (Fig.~\ref{fig:fig3}a2). This compares well with the findings of Xu \etal \cite{xu2019quantitative} who found the dominant presence of hydrophobic residues in both positions of bitter di-peptides. Observe that we derived the patterns in Fig.~\ref{fig:fig3} using the \emph{surrogate measures} for tastes that we defined; we did not have labels or bitterness values for the combinatorially constructed library peptides. The similarity between the second best bitter peptide pattern from our analysis and Xu \etal's \cite{xu2019quantitative} findings computed with 12 amino acid descriptors on labeled bitter di-peptides dataset demonstrates the \emph{strength of the surrogate measure} we defined. 

From the $N=3$ library onward, we started getting more robust predictions across the training sets. We found that the best predicted bitter and umami patterns for $N=3$ library are `HPP' and `$--$P' (with `HPH' and `$-$PP' as close second best patterns), respectively (Fig.~\ref{fig:fig3}a3,b3). Interestingly, the close second-best bitter pattern, `HPH', matches well with Xu \etal's result \cite{xu2019quantitative} who found hydrophobicity of the C-terminal residue and electronic properties of the second residue are important for bitterness in tri-peptides. The bitter pattern we got from the $N=4$ library, `HPPP', (Fig.~\ref{fig:fig3}a4) also compares well at residue positions one, three, and four with Xu \etal's findings for tetra-peptides. However, for the second position, they found that hydrophobicity plays a role; we got a polar residue in our predicted pattern.  

We could not find systematic sequential residue type analysis for bitter peptides with more than four residues and umami peptides with more than three residues. As a consequence, the longer patterns we found---\{HPPPP, HPPPPH, HPPPPHH\} for bitter (Fig.~\ref{fig:fig3}a5--a7) and \{$--$PP, $--$PPP, $--$PPPH, and $--$PPPHH\} for umami peptides (Fig.~\ref{fig:fig3}b4--b7)---can provide useful templates for exploring new bitter and umami peptides. Note that different residue types closely compete for the sixth and seventh residue position of the $N=6,7$ umami patterns. Also, the dominant residue type at each sequence position mostly remains conserved across the libraries from $N=2$ and up. The presence of several predictions from these libraries naturally leads to the question: which library and its predicted bitter and umami patterns do we select if we need to pick one `best' pattern for each taste?

\begin{figure}
    \begin{center}
        \includegraphics[width=\columnwidth]{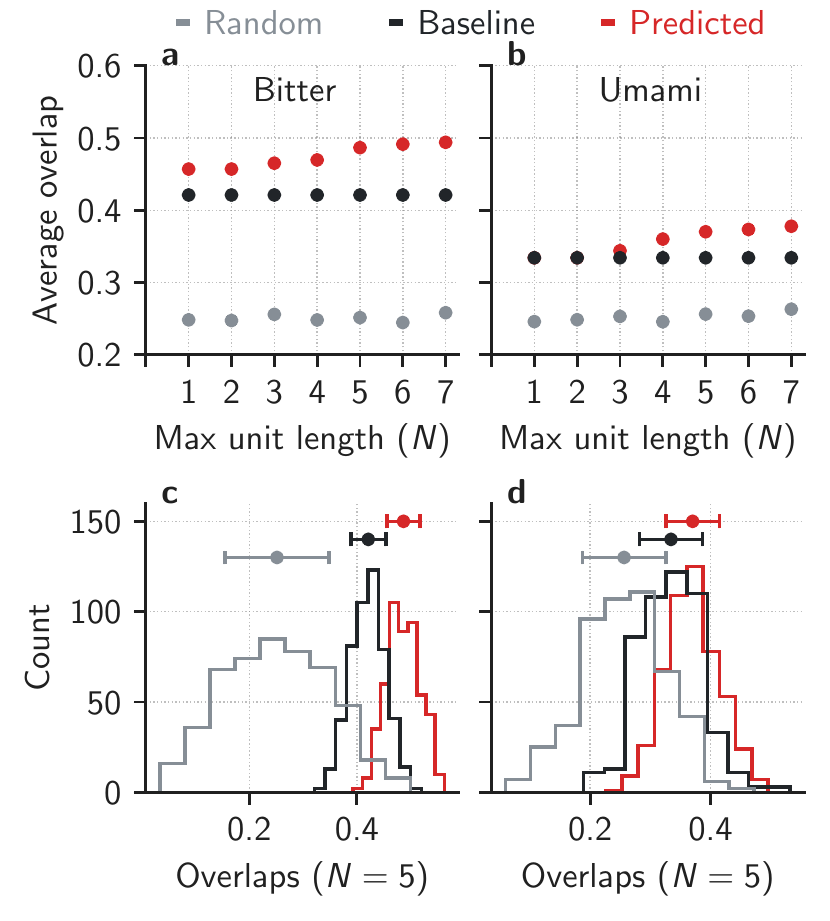}
        \caption{Increasing the maximum length, $N$, of the residue patterns---\ie allowing for more complexity in pattern libraries---doesn't lead to a better pattern above $N=5$. Panels \textbf{a} and \textbf{b} show average overlaps of the predicted bitter and umami patterns from each library (red dots), the baseline bitter (`HH$\cdots$') and umami (`$--\cdots$') patterns (black dots), and a randomly generated residue pattern (gray dots) with bitter and umami peptides from 500 test sets, respectively. The standard errors of the means are smaller than the dots' sizes. Panels \textbf{c} and \textbf{d} show histograms of overlaps of the $N=5$ library's predicted bitter (`HPPPP') and umami (`$--$PPP') patterns with bitter and umami peptides from the test sets, respectively. The ordinates denote the number of test sets with overlaps in a certain range. Horizontal lines above the histograms indicate the means and standard deviations of the distributions.}
        \label{fig:fig4}
    \end{center}
\end{figure}

\subsection{Selecting the minimal peptide pattern}
\label{sec:selecting}

To answer this question, we computed the average overlaps of the predicted bitter and umami patterns, which we found using the training sets, with the corresponding test sets' bitter and umami peptides, for each of the seven libraries (Fig.~\ref{fig:fig4}a and b, shown as red dots). We also computed the average overlaps of the baseline bitter (`HH$\cdots$') and the baseline umami (`$--\cdots$') patterns (black dots, Fig.~\ref{fig:fig4}a and b) and the average overlaps of a peptide with random-residues (gray dots, Fig.~\ref{fig:fig4}a and b). All averages were computed with the test sets' bitter and umami peptides. The standard errors of means of the averages are smaller than the sizes of the dots.

For bitter peptides, the predicted patterns for the smallest two libraries ($N=1,2$) are entirely made of hydrophilic residues (Fig.~\ref{fig:fig3}a1, a2). The average overlaps for those predicted patterns (red dots, Fig.~\ref{fig:fig4}a) are larger than the all-hydrophobic baseline pattern (black dots, Fig.~\ref{fig:fig4}a). This counterintuitive result, however, is an artifact of having a large number of hydrophilic residues in the bitter peptide dataset, as discussed earlier. For umami peptides, the smallest predicted patterns ($N=1,2$) are entirely made of negative residues (Fig.~\ref{fig:fig3}b1, b2). Because we have considered an all-negative residue as our umami baseline, the overlaps of the predicted patterns and baseline patterns with the test sets' umami peptides match (overlapped red and black dots, Fig.~\ref{fig:fig4}b). For both bitter and umami peptides, with increasing $N$, the overlaps increase until $N=5$; then they mostly plateau. From this observation, we chose the $N=5$ library as the minimal library that is large enough to have enough coarse-grained peptide patterns so that it neither underfits the data nor has more peptide patterns than necessary, given the model complexity.

To demonstrate how the $N=5$ peptide library overlaps with peptides from the test sets, we computed the histograms of the average overlaps of the $N=5$ library's predicted patterns---`HPPPP' for bitter and `$--$PPP' for umami peptides---with bitter and umami peptides from the test sets (Fig.~\ref{fig:fig4}c, d). The predicted patterns (in red) clearly are improvements over the baseline patterns (in black) and randomly-chosen patterns (in gray). The improvements are more pronounced for the bitter pattern compared to the umami pattern. This analysis demonstrates the accuracy of our method, which is designed to identify generic sequence patterns rather than predicting a property or classifying a new peptide. Taken together, these observations imply that the predicted bitter and umami patterns can act as promising design templates for bitter and umami peptides.

For bitter and umami peptides, our analysis offers a set of coarse-grained residue patterns that are possibly linked to the peptides' bitter and umami tastes. The predicted bitter patterns can be useful, for example, in finding short, possibly bitter-causing, patterns in longer proteins. As test cases, we considered two proteins that are associated with bitter taste: Patatin-T5 (UniProt ID: P15478) \cite{spelbrink2015potato, Patatin-T5} and Legumin A (UniProt ID: P02857) \cite{real2019enzymatic, Legumin-A}. We first converted the primary sequence of these proteins to a sequence of coarse-grained residues (see the supplementary information for the full primary and coarse-grained sequences) and then searched for the predicted pattern from $N=5$ library: `HPPPP'. The search resulted in eight five-residue-long sequence segments in Patatin-T5: \{ATTNS (2), ATTSS (16), IGGTS (73), ITTPN (86), FQSSG (114), ATNTS (200), LGTGT (258), LTGTT (324)\}. The numbers in the parentheses denote the segments' first residues' positions in the protein's primary structures. For Legumin-A, we found five such short sequences: \{IQQGN (93), IGPSS (347), CNGNT (418), VPQNY (436), AGTSS (468)\}. The analysis provides a possible experimental way to determine where to cleave the protein to decrease its bitterness. Accordingly, it may be useful to study these short sequences further in experiments.

\section{Conclusions}
In this study, we aim to build a simple and interpretable model that identifies generic residue patterns that are prevalent in bitter and umami peptides and, possibly, evoke those tastes. While a complex quantitative structure--activity relationship (QSAR) model or a machine learning (ML) model can offer accurate predictions \emph{given} a peptide sequence, their low interpretability and scarcity of peptide taste data limit their use as a tool for identifying generic patterns. Our model, instead of competing with QSAR and ML models in terms of accuracy, aims to complement these by providing a way to identify such patterns hidden in bitter or umami peptides and proteins.

By coarse-graining the twenty canonical amino acid residues into four physicochemically relevant classes---hydrophobic (H), hydrophilic (P), positively charged (+), and negatively charged ($-$)---we drastically reduced the dimensionality of the peptide sequence space. With these coarse-grained residues, we systematically built seven increasingly larger, more complex, combinatorial libraries of peptides. We compiled and coarse-grained a library of bitter and umami peptides from the literature. We used a sequence overlap index \cite{schilling2019sequence} $I$ (Eq.~\ref{eq:overlap-index})  to compare library peptides with the bitter and umami peptides, at a coarse-grained level. Importantly, the overlap index allowed us to compute the average overlaps of a library peptide with bitter and umami peptides from the compiled dataset and define those average overlaps as surrogate measures for bitterness and umami-ness, respectively. For each of the seven peptide libraries, the best (\ie most overlapping) bitter and umami patterns provided us with the predicted patterns. We checked the robustness of the predicted patterns through 80\%--20\% train--test splitting, and we reported the patterns that we got after averaging over 500 such splits. By comparing the average overlap of the predicted patterns with test set peptides, we found the minimal $N=5$ library whose patterns---`HPPPP' for bitter and `$--$PPP' for umami peptides---have almost equal overlap compared to larger libraries. Further, we found that these predicted patterns represent the known bitter and umami peptides more accurately than baseline patterns---all hydrophobic residues for bitter peptides and all negatively charged residues for umami peptides---and peptides with randomly-chosen residues.

\begin{figure*}
    \begin{center}
        \includegraphics[width=\textwidth]{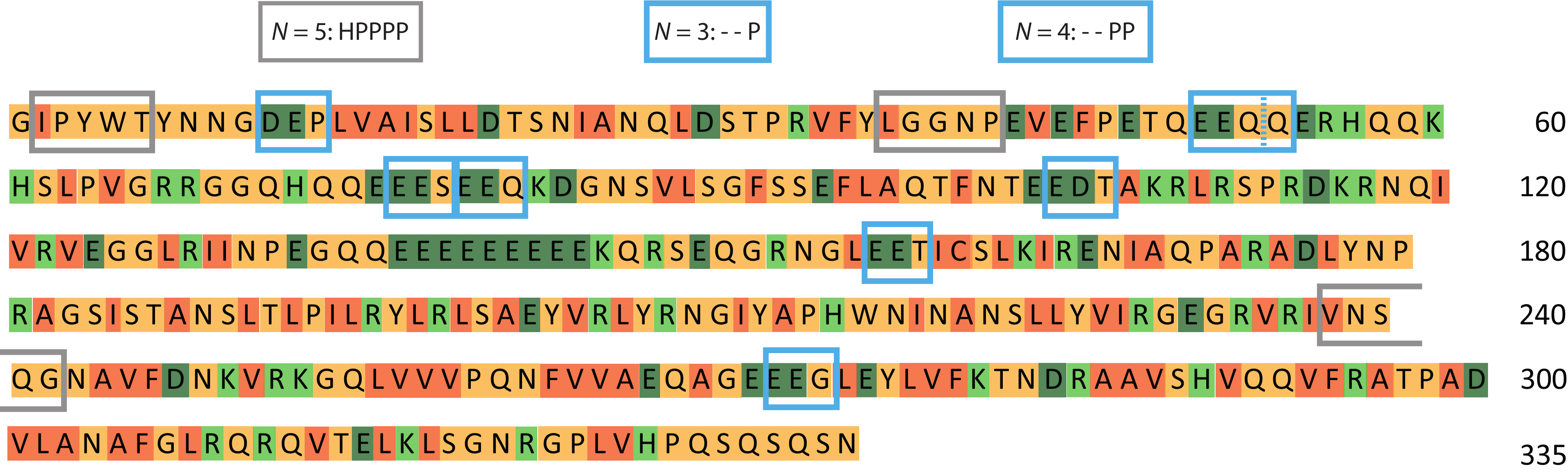}
        \caption{Illustration of the method along the protein Legumin B (Uniprot ID: P16078), which is present in broad beans (\textit{Vicia faba}). The primary structure is colored according to our scheme. Possible bitter and umami peptides are framed in grey- and blue-colored boxes, respectively.}
        \label{fig:fig5}
    \end{center}
\end{figure*}

The potential of the proposed coarse-grained peptide pattern search is demonstrated in Fig.~\ref{fig:fig5}. The amino acids of the primary structure of Legumin B (Uniprot ID: P16078) \cite{Legumin-B}, a storage protein from broad beans (\textit{Vicia faba)}, are colored according to the scheme from Fig.~\ref{fig:fig3}. Employing the results from Fig.~\ref{fig:fig3}, a search for the pattern HPPPP, the $N=5$ pattern with bitter taste, provides three peptides: IPYWT (2), LGGNP (38), and VNSQG (238) (the numbers denote the peptides' positions in the sequence). They are marked with grey boxes in Fig.~\ref{fig:fig5}. The umami potential of this protein is very high: a complete hydrolysis provides 38 glutamic and 10 aspartic acids---they constitute more than 14\% of the protein's total amino acid content. While there are no matches for the $N=5$ umami pattern `$--$PPP' in this particular protein, a search for the $N=4$ pattern `$--$PP' returns one peptide, EEQQ (51), and the $N=3$ umami pattern `$--$P' returns seven peptides: DEP (11), EEQ (51), EES (76), EEQ (79), EDT (104), EET (157), and EEG (271). All of them are marked with blue boxes in Fig.~\ref{fig:fig5}. Note that short bitter and umami pattern matches from the reversed and cleaved primary structure of the protein may also have bitter and umami tastes. These predicted patterns can be used, for example, in experimentally designing new umami peptides, or in finding short bitter or umami segments in a long protein. Also, specially chosen or designed enzymes that can cleave proteins at defined peptide bonds may be appropriate for guiding taste profiles of hydrolysates; they may then be used as new flavors of plant origin.

In conclusion, our coarse-grained peptide search provides a simple and quick screening for the taste potential of proteins. Such issues are becoming more important for plant-based meat surrogates. However, one point remains clear: many umami peptides, especially the longer ones, of animal origin are signatures of the specific proteins, and it remains difficult and seemingly impossible to find those in plant proteins.

\begin{acknowledgments}
    We acknowledge four open-source packages---Numpy \cite{harris2020array}, Matplotlib \cite{hunter2007matplotlib}, Scikit-learn \cite{pedregosa2011scikitlearn}, and Pandas \cite{mckinney2010data}---that were used in this work. A.D. acknowledges support by BiGmax, the Max Planck Society's Research Network on Big-Data-Driven Materials-Science. T.B. was partially supported by the Emmy Noether program of the Deutsche Forschungsgemeinschaft (DFG).
\end{acknowledgments}
\bibliography{ref}

\clearpage
\widetext
\begin{center}
\textbf{\large Supplementary information}
\end{center}

Here we have written out the complete original and coarse-grained primary structures of the two bitter proteins mentioned in the Sec.~\ref{sec:selecting}, \textit{Selecting the minimal peptide pattern}, of the main text. The coarse-grained residues are denoted as follows: `H' for hydrophobic, `P' for polar and hydrophilic, `+' for positively charged, and `-' for negatively charged.

\begin{itemize}
    \item {Patatin-T5}\cite{Patatin-T5}
    
    \textbf{FASTA entry}
    \begin{verbatim}
    >sp|P15478|PATT5_SOLTU Patatin-T5 OS=Solanum tuberosum OX=4113 PE=1 SV=1
    MATTNSFTILIFMILATTSSTFATLGEMVTVLSIDGGGIKGIIPATILEFLEGQLQEVDN
    NTDARLADYFDVIGGTSTGGLLTAMITTPNETNRPFAAAKDIVPFYFEHGPKIFQSSGSI
    FGPKYDGKYLMQVLQEKLGETRVHQALTEVAISSFDIKTNKPVIFTKSNLAKSPELDAKM
    YDICYSTAAAPTFFPPHYFATNTSNGDKYEFNLVDGAVATVDDPALLSISVATKLAQVDP
    KFASIKSLNYKQMLLLSLGTGTTSEFDKTYTAEETAKWGTARWMLVIQKMTSAASSYMTD
    YYLSTAFQALDSQNNYLRVQENALTGTTTELDDASEANMQLLVQVGEDLLKKSVSKDNPE
    TYEEALKRFAKLLSDRKKLRANKASY
          \end{verbatim}
          \textbf{Coarse-grained sequence}
          
          \begin{verbatim}
    HHPPPPHPHHHHHHHHPPPPPHHPHP-HHPHHPH-PPPH+PHHPHPHH-HH-PPHP-H-P
    PP-H+HH-PH-HHPPPPPPPHHPHHHPPPP-PP+PHHHH+-HHPHPH-+PP+HHPPPPPH
    HPP+P-P+PHHPHHP-+HP-P+H+PHHP-HHHPPH-H+PP+PHHHP+PPHH+PP-H-H+H
    P-HHPPPHHHPPHHPP+PHHPPPPPP-+P-HPHH-PHHHPH--PHHHPHPHHP+HHPH-P
    +HHPH+PHPP+PHHHHPHPPPPPP-H-+PPPH--PH+PPPH+PHHHHP+HPPHHPPPHP-
    PPHPPHHPHH-PPPPPH+HP-PHHPPPPP-H--HP-HPHPHHHPHP--HH++PHP+-PP-
    PP--HH++HH+HHP-+++H+HP+HPP
    \end{verbatim}

    
    \item {Legumin A}\cite{Legumin-A}
    
    \textbf{FASTA entry}
          \begin{verbatim}
    >sp|P02857|LEGA_PEA Legumin A OS=Pisum sativum OX=3888 GN=LEGA PE=1 SV=1
    MAKLLALSLSFCFLLLGGCFALREQPQQNECQLERLDALEPDNRIESEGGLIETWNPNNK
    QFRCAGVALSRATLQRNALRRPYYSNAPQEIFIQQGNGYFGMVFPGCPETFEEPQESEQG
    EGRRYRDRHQKVNRFREGDIIAVPTGIVFWMYNDQDTPVIAVSLTDIRSSNNQLDQMPRR
    FYLAGNHEQEFLQYQHQQGGKQEQENEGNNIFSGFKRDYLEDAFNVNRHIVDRLQGRNED
    EEKGAIVKVKGGLSIISPPEKQARHQRGSRQEEDEDEEKQPRHQRGSRQEEEEDEDEERQ
    PRHQRRRGEEEEEDKKERGGSQKGKSRRQGDNGLEETVCTAKLRLNIGPSSSPDIYNPEA
    GRIKTVTSLDLPVLRWLKLSAEHGSLHKNAMFVPHYNLNANSIIYALKGRARLQVVNCNG
    NTVFDGELEAGRALTVPQNYAVAAKSLSDRFSYVAFKTNDRAGIARLAGTSSVINNLPLD
    VVAATFNLQRNEARQLKSNNPFKFLVPARESENRASA
        \end{verbatim}
        \textbf{Coarse-grained sequence}
        \begin{verbatim}
    HH+HHHHPHPHHHHHHPPHHHH+-PPPPP-HPH-+H-HH-P-P+H-P-PPHH-PPPPPP+
    PH+HHPHHHP+HPHP+PHH++PPPPPHPP-HHHPPPPPPHPHHHPPHP-PH--PP-P-PP
    -P++P+-++P+HP+H+-P-HHHHPPPHHHPHPP-P-PPHHHHPHP-H+PPPPPH-PHP++
    HPHHPP+-P-HHPPP+PPPP+P-P-P-PPPHHPPH++-PH--HHPHP++HH-+HPP+P--
    --+PHHH+H+PPHPHHPPP-+PH++P+PP+P-------+PP++P+PP+P---------+P
    P++P+++P------++-+PPPP+P+P++PP-PPH--PHHPH+H+HPHPPPPPP-HPPP-H
    P+H+PHPPH-HPHH+PH+HPH-+PPH++PHHHHP+PPHPHPPHHPHH+P+H+HPHHPHPP
    PPHH-P-H-HP+HHPHPPPPHHHH+PHP-+HPPHHH+PP-+HPHH+HHPPPPHHPPHPH-
    HHHHPHPHP+P-H+PH+PPPPH+HHHPH+-P-P+HPH
        \end{verbatim}

\end{itemize}
\end{document}